\documentclass[prb, twocolumn, nofootinbib]{revtex4-2}
\usepackage{amssymb}
\usepackage{amsmath}
\usepackage{graphicx}
\usepackage{bm}
\usepackage{xcolor}
\usepackage[unicode=true,colorlinks=true]{hyperref}
\hypersetup{  
     urlcolor = blue }

\begin{document}

\title{Electron density effect on spin-orbit interaction in [001] GaAs quantum wells }

\author{P.~S.~Alekseev}
\author{M.~O.~Nestoklon}
\affiliation{Ioffe Institute, St.~Petersburg 194021, Russia}

\begin{abstract}

 The spin-orbit interaction of two-dimensional (2D) electrons in semiconductor quantum wells gives rise to a varity of interesting transport and optical spin-dependent effects.  
  In the GaAs/AlGaAs type heterosystems, this interaction
  consists of the isotropic  Bychkov-Rashba term,
  which is absent in symmetric wells,  and the anisotropic Dresselhaus term,
  reflecting the lattice symmetry.
    It is well-known that the first term can be controlled by electric fields in the growth direction: external or internal, induced by a charge density of 2D electrons.
 In this work we reveal that the 2D electron charge can substantially affect also the Dresselhaus interaction in symmetric quantum wells.
  Within the one-band electron Hamiltonian
containing, together with the bulk Dresselhaus  interaction, the two
contributions to the Dresselhaus term from  the quantum well interfaces,
 we show that the  internal electric field from the  2D electron charge density
 can substantially renormalize
the anisotropic spin-orbit interaction of 2D electrons.
  This effect may be important in quantitative studies of spin-dependent phenomena in quantum wells.
\end{abstract}


\maketitle

\section{Introduction}
During the last three decades, extensive investigations of
the spin-orbit interaction of two-dimensional (2D) electrons in the
heterostructures based on the non-centrosymmetric cubic
semiconductors were performed  \cite{Winkler_book,Zutic04,bookIvchenko,Ganichev14}.
For the bulk samples from these material the general form of the
one-band effective mass electron Hamiltonian with the spin-orbit
terms 
was established
from a symmetry consideration~\cite{Dr,Ogg}. The magnitudes of these
terms can be derived from the many-band Kane {\bf k}$\cdot${\bf
p}-models~\cite{Ogg,Zaw,Pikus} or obtained in microscopic
calculations~\cite{Cardona88,Z5,Nestoklon12,Tarasenko15,Nestoklon16,Alekseev17}.

The spin-orbit interaction of 2D electrons in the heterostructures
grown from 
 the typical  III-V semiconductors consists of the two parts with
different symmetry: the Bychkov-Rashba term induced 
asymmetry of the structure in the growth direction
and the Dresselhaus term related to the spin-orbit interaction in the
crystal lattice. 

In addition to the conventional bulk
contribution to the anisotropic Dresselhaus interaction,
 there  also exist the two anisotropic interface contributions, determined by
the atomic structure of the interfaces of III-V semiconductors
\cite{Alekseev1,Alekseev2,Alekseev17}.
 For qualitative  explanation  of the experiments on spin-dependent phenomena
 in 2D electron gas in quantum wells, both the bulk and the interface contributions
  are important (see for example Refs.~\cite{Alekseev2},\cite{13}).

The effect of the sharp interfaces on the states in the III-V
semiconductor heterostructures is known since 1980s~\cite{Chang83,Schulman85,Chang85}.
Corresponding interface anisotropic terms were
introduced in the effective mass Hamiltonian for
holes to explain the nature of the anisotropic exchange splitting of
excitonic levels \cite{Aleiner92,Ivchenko93}.
Later, it has been noticed \cite{Vervoort99,Rossler}  that the
mixing of heavy and light holes at the interfaces also leads
to a spin-orbit interface anisotropic term in the one-band
electron Hamiltonian, being additional to the bulk Dresselhaus contribution.
 This term is one of the possible contributions to
the interface anisotropic spin-orbit interaction of 2D electrons in III-V
heterostructures, described within one-band Hamiltonian in Refs.~\cite{Alekseev1,Alekseev2,Alekseev17}.
In Refs.~\cite{Nestoklon12,Nestoklon16,Alekseev17,Erratum} the parameters of the interface contributions were extracted from the tight-binding calculations of quantum wells grown in [110] and [001] directions.
From the comparison, the values of the bulk and the interface parameters in the
one-band electron Hamiltonian have been extracted.
It was shown that the analytical one-band calculations perfectly reproduce the tight-binding numerical
calculations for different quantum wells.

In this work we study the effect of the internal electrostatic field induced by the charge density of 2D electrons on the spin-orbit coupling of 2D electrons in symmetric [100] GaAs quantum wells. 
We use the one-band model  \cite{Alekseev1,Alekseev2,Alekseev17} with account on  the bulk as well as the interface terms in the effective electron spin-dependent Hamiltonian. We find the dependence of the spin splitting of the energy spectrum of 2D electrons on the electron density by solving the self-consistent system of the Schrodinger and the Poisson equations for the wave functions in a well.
We show that the increase of the electron density substantially renormalizes
 the contributions  from the bulk as well as the interfaces
 to the anisotropic part of the spin splitting of the 2D electron spectrum.
  A raise of the contribution from the interfaces to the Dresselhaus 
  spin-orbit interaction stems from the 
  pushing out of the electron wave function 
  closer to the interfaces due to the Coulomb repulsion 
  of electrons in the well.
 We conclude that
  the effect of change density on the spin splittings can be important in
 a precise determination of the magnitudes of the bulk and the interface spin-orbit parameters
in GaAs type quantum wells.

\section{Structure design}
First, we discuss the possible structure design and the realistic values of their parameters for which the effect of the 2D electron charge on the spin splittng of the energy spectrum
 is expected to be substantial.

 We consider that the proper  GaAs/AlGaAs heterostructures are similar to standard structures designed
 to obtain an ultra-high mobility 2D electron gas (2DEG).
Let the barrier and the well layers be fabricated from the following typical compounds:
Ga$_{0.7}$Al$_{0.3}$As$/$GaAs$/$Ga$_{0.7}$Al$_{0.3}$.
 To reach  high electron densities in a quantum well (QW), the $\delta$-doping is usually used.
 To avoid the Rashba effect, the heterostructure should be symmetric in the growth direction.
 The GaAs/AlGaAs heterosystems of this type were
studied, for example,  in Ref.~\cite{Umansky09}. 
Such structures are usually optimized to have the maximum mobility that implies that the doping layers are located far from the QW. However, 
 in order to maximize the electron concentration of  the 2DEG inside the QW, the doping layers should be as close to the QW as possible. Below we consider the structure design with
  the $\delta$-doping layers located at the distance 20~nm from the QW, which still provides a sufficiently high mobility and a sufficiently  large electron concentration~\cite{Umansky97}.

It is natural to expect that the effect of the 2D electron charge on the spin splitting is large if the QWs are sufficiently wide. However, in wide QWs, if the 2D electron concentration is large, more than one electron subband can be populated. We do not discuss this case, as additional terms in spin-splitting electron Hamiltonian can appear due to the inter-subband mixing of spin-dependent wave functions(see e.g. \cite{Egues2015,Loss2008}).
 To determine the diapason of possible structure parameters when only lower subband is populated, we solve the system of the Poisson and the Sch\"odinger equations for electrons in QWs at different QW widths. We increase the Fermi level and find the concentration when the lower electron subbands start to populate.
 For thin QWs, there exists a maximum electron concentration when Fermi level reaches the conduction band in the barriers. This maximum concentration strongly depends on the distance between QW and the $\delta$-doping layers.

\begin{figure}
\includegraphics[width=\linewidth]{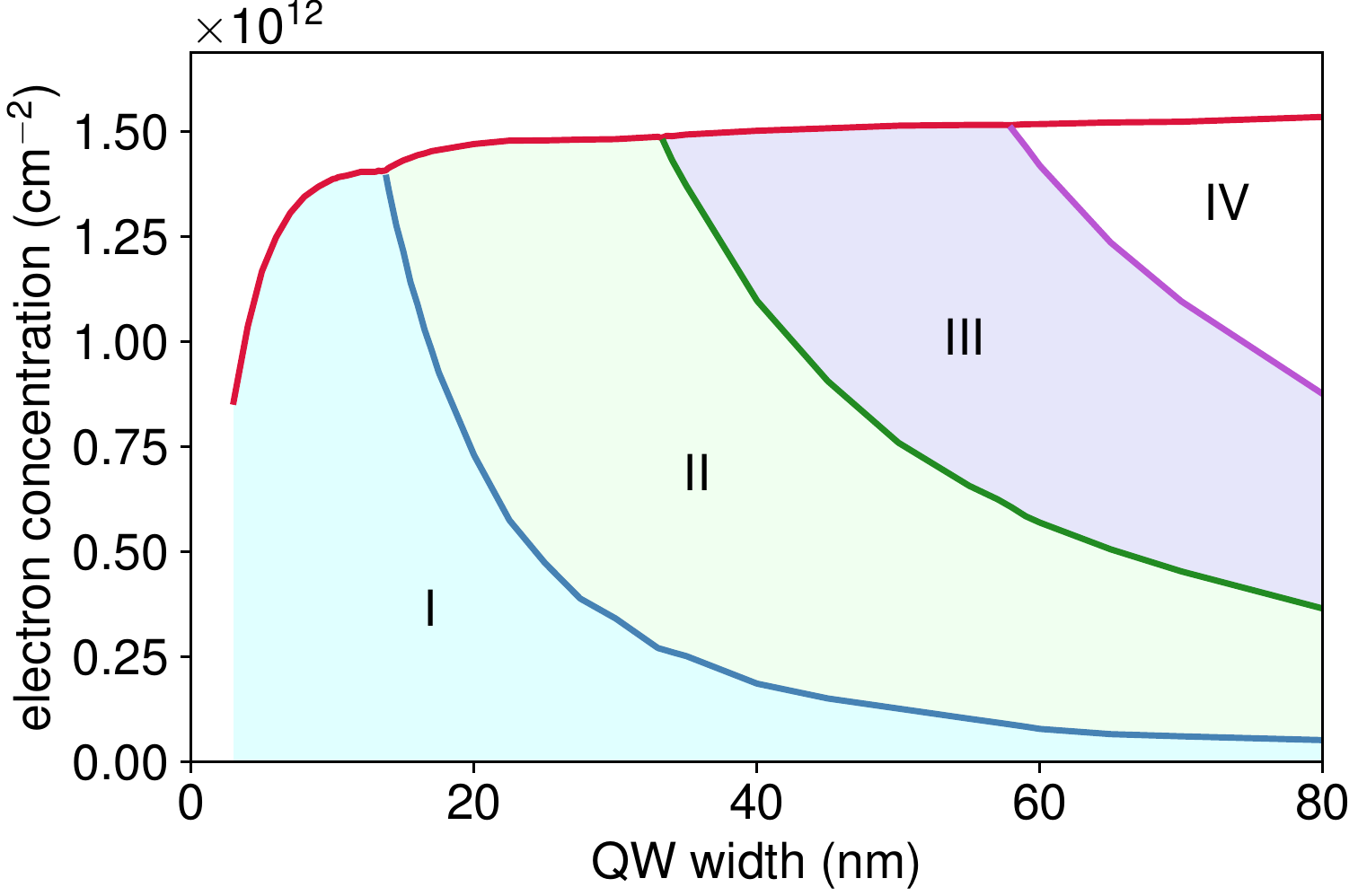}
 \caption{
     Regions of qualitatively different structure of 2DEG in QW as a function of QW width. Red line shows the maximum possible electron density in QW, blue line bounds the region of concentrations with only one electron subband populatred, green and violet lines bound the regions with two and three subbands populated respectively. Roman numbers $\mathrm{I}$,  $\mathrm{II}$, $\mathrm{III}$, and $\mathrm{IV}$ show the regions of the QW widths and the electron density where 2DEG populates 1, 2, 3, and more subbands, respectively.
}\label{fig:nmax}
\end{figure}

Results of solution of the system of the Poisson and the Schr{\"o}dinger equations for the electron gas in quantum well as a function of QW width are shown in Fig.~\ref{fig:nmax}. The charge inside QW is compensated by the $\delta$-doping which is modeled by the charge distributed in regions of the width of 2~\AA\  at the  distance of 20~nm from the QW boundaries. From Fig.~\ref{fig:nmax} it is seen that the QW in this structure can not be populated above $1.3\cdot10^{12}$~cm$^{-2}$. We expect that in real structures this value is somewhat smaller because in our estimations we do not consider the details of doping (donor activation energy is neglected). When comparing with experimental values from standard high-mobility structures design~\cite{Umansky97}, note that we assume symmetric $\delta$-doping which doubles maximum possible density as compared with asymmetric $\delta$-doping.

From Fig.~\ref{fig:nmax} it is clear that the most favorable simple QW design
for which the charge density inside the QW is maximal, being herewith distributed only 
in the first subband, corresponds to  the QW width around 13~nm.
 However, it remains to be  an open question whether the change of the electron density (relative to an almost empty well)  at the points of the interfaces is maximal at the same QW width, $\sim$13~nm, or it is maximal for larger QWs.

\section{One-band Hamiltonian approach to spin splittings of 2D electrons}
For quantum wells grown along the [001] direction from the zincblende
semiconductors, the bulk spin-orbit
term in the one-band electron Hamiltonian \cite{Pikus} has the form:
\begin{equation}
\label{Dr}
 \hat{H}_{BIA}=\frac{d}{dz} \gamma(z)\frac{d}{dz}
(k_x\hat{\sigma}_x - k_ y \hat{\sigma}_y)
 \:,
\end{equation}
 where $\gamma(z)$ is the bulk
spin-orbit parameter, which depends on the layer material.
In this formula we retained only the linear in the electron lateral
wavevector $ \bm{k} = (\, k_x , \, k_y \, ) $ contribution. The term~(\ref{Dr})
 is anisotropic relative to rotations in the
 $ \bm{k}$ plane.

 In quantum wells from zincblende
semiconductors with abrupt interfaces,
in addition to the bulk contribution~(\ref{Dr}),
 the anisotropic spin-orbit interaction
also contains the interface contributions~\cite{Rossler,Golub04,Nestoklon06,Nestoklon08,Alekseev1,Alekseev2,13,Dettwiler17}.
They have the same symmetry as the bulk term
\eqref{Dr}, but are proportional, instead of the operator $d^2/dz^2$,
to the delta-function and its derivative localized at the quantum
well interfaces \cite{Alekseev2,Alekseev1}:
\begin{equation}
\label{int}
\begin{split}
\hat{H}_{int}= & \sum \limits _{\nu=l,r } (\hat{H}_{int,0, \nu}+ \hat{H}_{int,1,\nu})
\:,
\\
\hat{H}_{int,0,\nu} = &  \zeta_{\nu}
\delta(z-z_{\nu}) (k_x\hat{\sigma}_x - k_ y \hat{\sigma}_y)
\:,
\\
\hat{H}_{int,1,\nu} =  & \xi_{\nu}
\delta'(z-z_{\nu}) (k_x\hat{\sigma}_x - k_ y \hat{\sigma}_y)
\:.
\end{split}
\end{equation}
The parameters  $\zeta_{l,r}$ and  $\xi_{l,r}$ are determined by the
structure of the chemical bonds of the atoms on the interfaces.
Equations (\ref{int})~imply the continuity of the wave function
derivative at the interfaces of the quantum well.

Comparison with the tight-binding calculations for
Ga$_{0.7}$Al$_{0.3}$As$/$GaAs$/$Ga$_{0.7}$Al$_{0.3}$ quantum well
gives the following values of the interface parameters and
of the bulk cubic-in-k aplitting: $\xi=1.5$~eV$\cdot
${\AA}$ ^3$, $\zeta =0.124$~eV$\cdot ${\AA}$ ^2$,
$\gamma=\gamma(\text{GaAs})=\gamma_w = 23$~eV$\cdot ${\AA}$ ^3$,
$\gamma_b=\gamma(\text{Al\textsubscript{0.3}Ga\textsubscript{0.7}As})=0.7\gamma$ (see Ref.~\cite{Alekseev17} for the details).

In the following,
 we consider a quantum well with a rectangular heteropotential $U_0(z)$. The function $\gamma(z)$ in
this structure is a step-like function with the two different values
in the well and in the barrier layers:
\[
\gamma(z)=
 \left|
 \begin{array}{l}
\gamma_b \:, \;\; z<-a/2 \, ,\; z>a/2
 \\
 \gamma_w \:, \;\; -a/2<z<a/2
 \end{array}
 \right.
 \:.
\]
In the absence of an electric field $E_z$,
the symmetry of the quantum well is $D_{2d}$ that
leads to the relations between the coefficients $\zeta_{\nu}$ and
$\xi_{\nu}$ for the left and right interfaces:
$\zeta_{l}=\zeta_{r}$, $\xi_{l}=-\xi_{r}$.

\begin{figure}
\includegraphics[width=\linewidth]{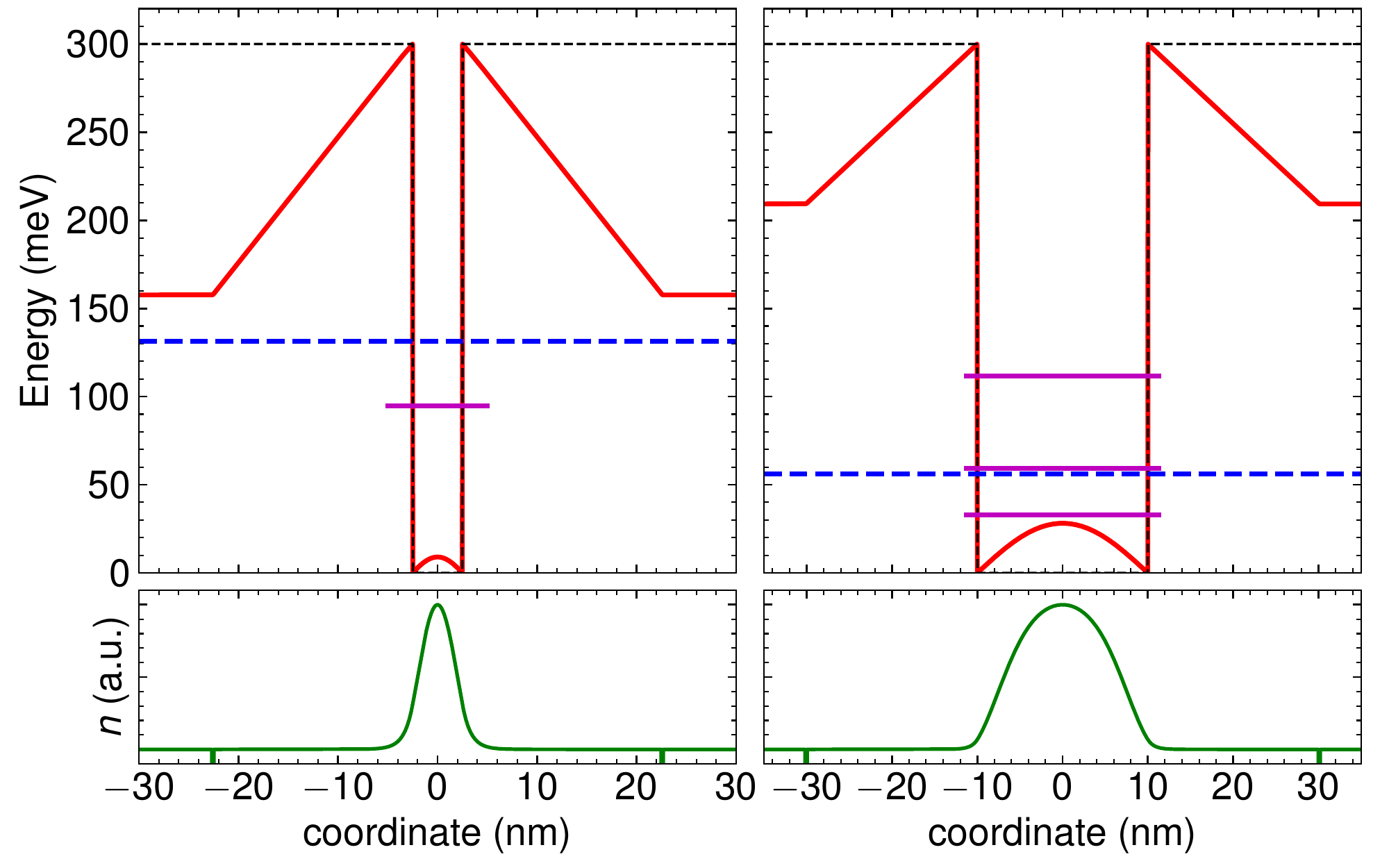}
    \caption{Profiles of electrostatic potential (upper panels) and electron density (lower panels) for two QWs with large electron density: QW width 5~nm and total density $n_{\mathrm{tot}}=1.02\cdot10^{12}$~cm$^{-2}$ (left panels) and QW width 20~nm and total density $n_{\mathrm{tot}}=6.45\cdot10^{11}$~cm$^{-2}$ (right panels). Profile of uncharged QW is shown in thin dashed line, the electrostatic profile of charged QW $U(z)$ is shown in red line. Fermi level $E_F$ is shown in blue dashed horizontal line.
}\label{fig:U_n_psi}
\end{figure}

The resulting anisotropic spin-orbit Dresselhaus Hamiltonian of 2D electrons
 containing the bulk and
the interface contributions (\ref{Dr}) and (\ref{int}) and
averaged by a wave function $u(z)=u_i(z)$ of a space-quantized level $E_i$ ($i=1,2,...$)
is:
\begin{equation}
 \hat{H}_{SO}
=
\beta(k_x\hat{\sigma}_x - k_ y \hat{\sigma}_y)
\:,
\end{equation}
where the spin splitting parameter $\beta$ consist of the three contributions:
\begin{equation}
\label{eq:betas}
\begin{split}
 \beta & = \beta_{b}+ \beta_{int,0} +\beta_{int,1}   \:,
 \\
 \beta_{b} & =  -  \int _{-\infty}^{\infty} dz \: \gamma (z) [u'(z)]^2  \:,
\\
 \beta_{int,0} & = \zeta \, [u(-a/2)^2+u(a/2)^2] \:,
\\
  \beta_{int,1} & = \xi \, \{[u(-a/2)^2]'-[u(a/2)^2]'\}  \:.
\end{split}
\end{equation}

The potential energy of an electron in a rectangular quantum
well with a substantial 2D electron charge density is:
\begin{equation}
\label{true}
 U(z) =
 e \Phi(z) + \left|
\begin{array}{l}
0, \; \;  -a/2  < z  <  a/2
\\
U_0, \; \;  z < -a/2 , \; z  >  a/2
\end{array}
\right.
\:,
\end{equation}
where $\Phi(z)$ is the electrostatic potential induced by the 2D electron charge density.
The potential $\Phi(z)$  is to be found from the solution of
the Poisson equation:
\begin{equation}\label{eq:Poiss}
    \frac{\partial^2}{\partial z^2}\Phi(z) = -\frac{4\pi e^2}{\varepsilon} \left[n(z)+n_{\delta}(z)\right]\:.
\end{equation}
Here we neglect the exchange contribution (see discussion in Ref.~\cite{Alekseev2}). 
The density of charge corresponding to $\delta$-doping is $n_{\delta}(z)$ and the
electron density $n(z)$ is found from
 the electron wave functions taking into account the electron  charge
 in the  QW:
\begin{equation}\label{eq:n}
 n(z) = \sum_{i
 } n_{i
 } \left| u_{i
 }(z)  \right|^2
 \:,
\end{equation}
where $i$-th subband electron density is:
\begin{equation}\label{eq:subband_DOS}
n_{i
%
} = \frac{m^*}{\pi\hbar^2} \Theta(E_F-E_{i
})\,,
\end{equation}
and $\Theta(E)$ is the Heaviside step function. In Eq.~\eqref{eq:subband_DOS} we neglect the finite temperature as we are interested in the case when the distance between electron levels
is large as compared with the temperature.
 Electron wave functions
 are obtained from the numerical solution of Sch\"odinger equation:
\begin{equation}\label{eq:Schr}
    \left[ -\frac{\hbar^2}{2m}\frac{\partial^2}{\partial z^2} + U(z) \right] u_{i
  }(z)
  = E_{i
  } u_{i
  }(z)\,.
\end{equation}

  In real structures, the doping density $n_{\delta}$ is defined by the structure design and Fermi level $E_F$ is found from the system of equations (\ref{eq:Poiss},\ref{eq:Schr}) under electroneutrality condition $n_{\mathrm{tot}}= \int n(z) = -\int n_{\delta}(z)$. However, this approach involves finding roots of inexplicit non-linear equations. Technically, it is more convenient to construct the solutions for a set of structures with different 2D electron density starting from the structure with zero density and  then increasing  the Fermi energy $E_F$ in small steps. Then, the doping density $n_{\delta}$ is straightforwardly found from the electroneutrality condition and doping profile. Provided choosing a sufficiently small step in Fermi energy, all the values (the energies, the wave functions, and the spin splitting constants) may be found as functions of electron density relatively fast. At each value of $E_F$, the
system of equations (\ref{eq:Poiss}) and (\ref{eq:Schr}) is solved iteratively until the convergence
 of the functions  $\Phi^{(j)}(z)$ and $u^{(j)}_i(z)$ at the successive iteration steps $j$ and $j+1$
  is achieved. To stabilize convergence, at each $j$-th step the electron density $n^{(j)}(z)$ is the weighted average between the density at previous step and the solution of Eq.~\eqref{eq:Schr} calculated from the density at previous step.

The profiles of electrostatic potential and electron density is shown in Fig.~\ref{fig:U_n_psi} for the two QWs: 
for a relatively narrow  5~nm  QW  with 
the density $n_{\mathrm{tot}}=1.02\cdot10^{12}$~cm$^{-2}$ and for a relatively wide QW 20~nm with 
the density $n_{\mathrm{tot}}=6.45\cdot10^{11}$~cm$^{-2}$.

\section{Results}
\begin{figure}
\includegraphics[width=\linewidth]{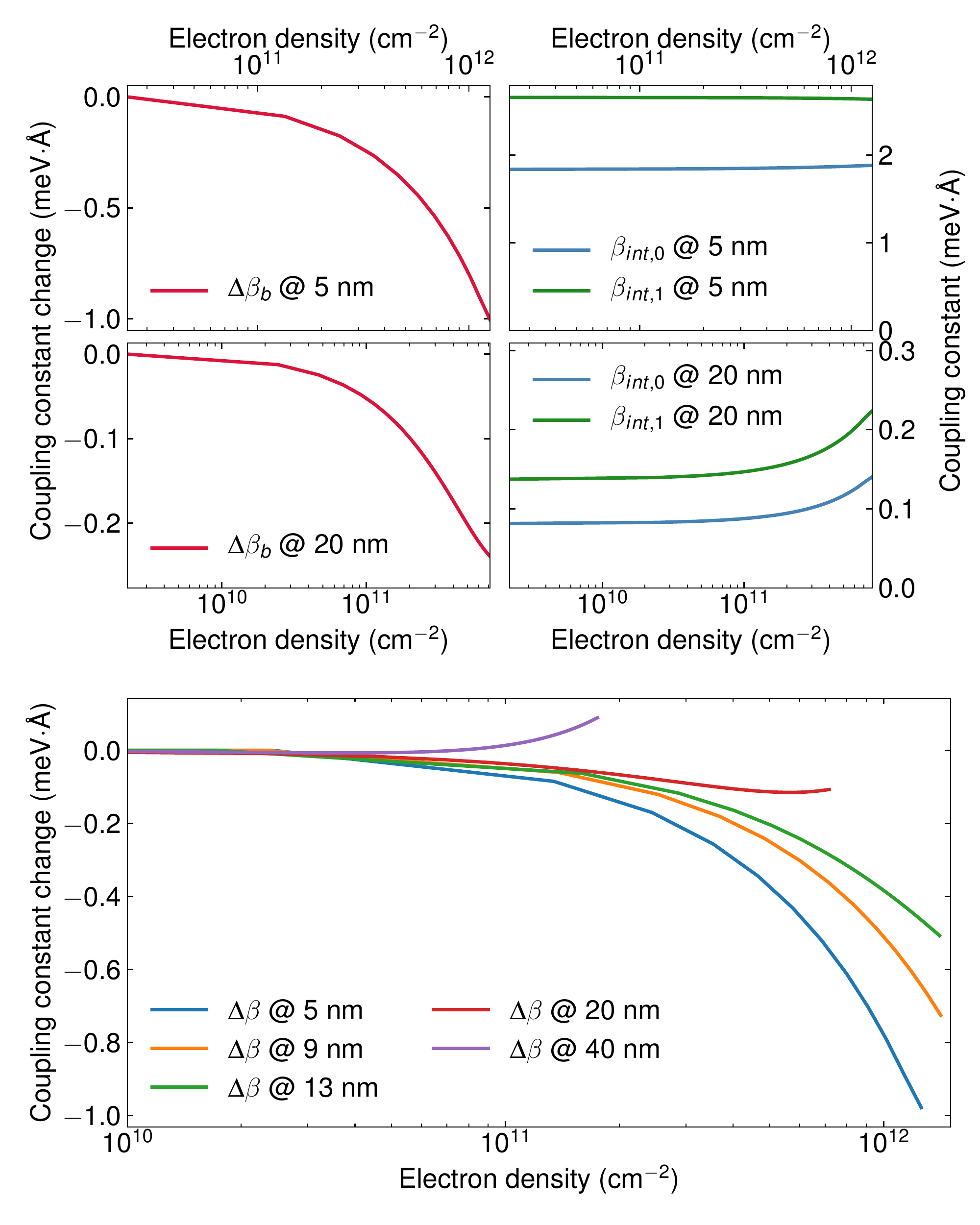}
 \caption{Bulk and interface contributions to the Dresselhaus constant $\beta$ as a function of electron density for quantum well width 5~nm (upper row) and 20~nm(second row).
 The values of the bulk contribution $\beta_b$
 to the constant $\beta$ at zero 2D electron density, $n_{\mathrm{tot}}=0$,
 are 19~meV$\cdot ${\AA}  for the 5~nm well and 3.8~meV$\cdot ${\AA}   for 20~nm well.
    Bottom panel shows the change of the total Dresselhaus constant $\beta(n_{\mathrm{tot}})-\beta(n_{\mathrm{tot}}=0)$ as a function of electron density for few selected QW widths.
}\label{fig:beta_vs_n}
\end{figure}

From the calculated wave functions in charged quantum well $u_0(z)$, we calculate the spin orbit constants $\beta_{b}$,  $\beta_{int,0}$, and $\beta_{int,1}$ using Eqs.~\eqref{eq:betas}.
The values of these constants  as a function of electron density in the QW for few typical QW widths are presented in Fig.~\ref{fig:beta_vs_n}. 

   In symmetric GaAs/AlGaAs QWs with small electron population, the interface terms $\beta_{int,0/1}$ are 
   comparable  (about 30 percent) with bulk contribution $\beta_b$ when the QW width is relatively small, 
   below $\sim$5~nm \cite{Alekseev17}. 
However, our current calculations show
 that  for such thin QWs the change of the interface contribution with the change of the electron density is small.
     Quite the contrary, our calculations also demonstrate that the effect of 2D electron charge 
      on the QW potential profile 
   is much more pronounced in large QWs.

Indeed, from Fig.~\ref{fig:beta_vs_n} it is clear that, at substantial 2D electron populations,   for the wide QW with $a=20$~nm the value of the interface contribution parameters 
  $\beta_{int,0/1}$ change significantly with changing the 2D electron density $n_{\mathrm{tot}}$, unlike their
almost independent on $n_{\mathrm{tot}}$ behavior
 in the narrow well  with $a=5$~nm. Such change of $\beta_{int,0/1}(n_{\mathrm{tot}})$, approximately two times with changing of $n$ from zero to maximum possible value,  is of the same order of magnitude 
 as the change of the bulk contribution to the Dresselhaus interaction,  
 $\beta_b(n_{\mathrm{tot}}) -\beta_b(n_{\mathrm{tot}}=0) $ with changing of $n_{\mathrm{tot}}$ up to maximum possible value. 
 Note that the main part of the bulk spin-orbit parameter  $\beta_b(n_{\mathrm{tot}}=0) $ is still much larger than interface contributions $\beta_{int,0/1}(n_{\mathrm{tot}})$  for the all studied QWs.

In lower panel of Fig.~\ref{fig:beta_vs_n} we present the numerical results for the change of the total Dresselhaus constant $\beta = \beta_{b} + \beta_{int,0}+\beta_{int,1}$, which is defined as 
\begin{equation}
    \Delta\beta(n_{\mathrm{tot}}) = \beta(n_{\mathrm{tot}})-\beta(n_{\mathrm{tot}}=0)\,,
\end{equation} as a function of electron density in QW.
Note that the spin splitting constant can exhibit different types of dependencies on the electron density $n_{\mathrm{tot}}$ for different width of the quantum wells. For thin quantum wells, the interface contribution is relatively large, but it does not change with the electron density. In such QWs, the change of the spin splitting constant $\beta$ is defined by the change of bulk contribution to the spin splitting $\beta_{b}$. When the QW is sufficiently large, the change of $\beta$ is defined by the interface constribution. However, in wide GaAlAs QWs the interface contribution is small compared with bulk contribution. Thus, the detailed calculations show that there are two distinct regimes of dependence of spin splitting constant on the electron concentration. For QW widths below 20nm it is mainly due to the bulk contribution, whereas for wider QWs it is due to the interface one.

The increase in the interface contribution to the Dresselhaus interaction has the following reason.
Due to the Coulomb repulsion,
2D electrons in wide quantum wells are ``pushed'' from the well center
to the regions near the interfaces
 and the relative significance of the interface spin-orbit terms 
 substantially increases, depending  on the 2D electron density.
Due to the difference in the bulk Dresselhaus constants $\gamma_b$ and $\gamma_w$ 
in the materials of the barrier and of the well, the bulk contribution to the 
anisotropic spin-orbit interaction decreases with the increase of~$n$.


\section{Conclusion}
We have shown that in GaAs/AlGaAs quantum wells with sufficiently high electron densities, at which  the 2D electron charge significantly affects the profile of the heterostructure potential $U(z)$, the
 interface as well as the bulk contributions to the anisotropic Dresselhaus spin-orbit interaction
noticably change as a function of electron density.
This effect should be important in quantitative studies of spin-orbit effects in these systems.

\section*{Acknowledgments}
Analysis of the heterostructure design, calculations of the electron wave functions and the spin-orbit constants, described in section II and in a part of section IV, was supported by the Russian Science Foundation (Grant 17-12-01265) and the Foundation for advancement of theoretical physics and mathematics ``BASIS''.

\bibliography{IIA}

\end{document}